\documentclass[12pt,preprint]{aastex}
\begin{document}

\title{The Identification of New Stellar Groupings in the M81 Debris Field\altaffilmark{1,}\altaffilmark{2,}\altaffilmark{3}}

\author{T. J. Davidge}

\affil{Herzberg Institute of Astrophysics,
\\National Research Council of Canada, 5071 West Saanich Road,
\\Victoria, B.C. Canada V9E 2E7\\ {\it email: tim.davidge@nrc.ca}}

\altaffiltext{1}{Based on observations obtained with the
MegaPrime/MegaCam, a joint project of the CFHT and CEA/DAPNIA,
at the Canada-France-Hawaii Telescope (CFHT), which is operated by
the National Research Council (NRC) of Canada, the Institut National des
Sciences de l'Univers of the Centre National de la Recherche
Scientifique (CNRS) of France, and the University of Hawaii.}

\altaffiltext{2}{This research has made use of the NASA/IPAC Extragalactic Database (NED), 
which is operated by the Jet Propulsion Laboratory, California Institute of Technology, 
under contract with the National Aeronautics and Space Administration.}

\altaffiltext{3}{This research used the facilities of the Canadian Astronomy Data Centre, operated by the National Research Council of Canada with the support of the Canadian Space Agency.}

\begin{abstract}

	The stellar content of the debris field around M81 is investigated using deep 
images obtained with MegaCam on the Canada-France-Hawaii Telescope. 
Three new concentrations of bright main sequence stars and red supergiants are identified 
within $\sim 20$ kpc of M81. These systems have integrated brightnesses M$_V \sim -11$ and 
surface brightnesses $\mu_V \sim 27 - 28$ mag arcsec$^{-2}$. The main 
sequence turn-offs are consistent with an age of $\sim 30$ Myr, while the presence 
of a well-developed red supergiant plume indicates that their stellar content is not 
coeval, as they also formed stars $\sim 100$ Myr in the past. The photometric 
properties of the red supergiants indicate that the young stars in these 
systems have metallicities that are comparable to those of other objects in the debris field. 
The HI density near these groupings comfortably exceeds the threshold required to trigger 
star formation. Based on the close proximity of these objects to M81, coupled with 
their extended sizes and low inferred masses, it is argued that they will not 
be long-lived structures, but will probably dissipate within the next $\sim 1$ Gyr. 

\end{abstract}


\section{INTRODUCTION}

	The galaxies in the local universe have not evolved in isolation; rather, their 
properties have almost certainly been shaped by interactions with other systems. 
The currently accepted galaxy formation paradigm is that the large galaxies in the 
local universe were assembled through the accretion of smaller systems. The accretion of 
galactic building blocks was more frequent during early epochs, and simulations suggest 
that the bulk of the Galaxy was in place $\sim 9$ Gyr ago (e.g. Bullock \& Johnston 2005).

	While the frequency of mergers has dropped with time, the merger rate may have 
remained high enough that most nearby disk galaxies still experienced significant mergers 
during the past few Gyr (e.g. Hammer et al. 2005).
That most spiral galaxies experienced cosmologically recent mergers challenges 
the classical notions that disks may be too fragile to survive such encounters, and that 
the inevitable consequence of a major merger is the formation of a pressure-supported system. 
However, recent simulations suggest that gas-rich disks may form (or re-form) after major 
mergers involving gas-rich progenitors (e.g. Brook et al. 2007; Governato et al. 
2007; Robertson et al. 2006; Springel \& Hernquist 2005). If such events occurred 
during relatively recent epochs then the specific angular momenta of disks 
would be boosted long after the earliest epochs of galaxy assemby. 
This could explain the difference between disk sizes predicted by CDM models and 
those that are observed (Hammer et al. 2007), and the tendency for 
disk sizes to increase with decreasing redshift (Trujillo et al. 2006).

	Hammer et al. (2007) argue that the nearest external spiral galaxy, M31, is a 
`typical' disk galaxy, in that its disk radius, specific angular momentum, and 
metallicity have been influenced by comparatively recent interactions/mergers. 
Indeed, there are a number of signatures of cosmologically recent interactions 
between M31 and companion galaxies. These include (1) a stellar stream (Ibata et al. 2001), 
(2) distinct sub-systems in the M31 globular cluster system, 
that may be indicative of the assimilation of a globular cluster system from a companion 
(e.g. Beasley et al. 2005), (3) a pervasive moderately metal-rich intermediate age 
extraplanar population that might be debris from disrupted satellites or an 
artifact of an interaction that stirred the disk of M31 
(Brown et al. 2006), and (4) the distribution of interstellar material 
in the disk (Gordon et al. 2006). 

	Studies of nearby spiral galaxies that are experiencing on-going interactions 
will provide insight into the role that galaxy-galaxy interactions 
have played in shaping the properties of systems like M31, and 
M81 is an obvious target for such a study. Yun, Ho, \& Lo (1994) find HI streams 
linking M81, M82, and NGC 3077, as well as isolated HI clouds, all of which are thought to 
have formed in response to an encounter between M81 and M82 a few hundred Myr in the past. 
Many of the areas that contain neutral hydrogen also contain stars (Sun et al. 
2005), suggesting that more than gas may have been stripped from M81 and/or M82.

	The pedigree of objects like Ho IX (DDO 66) and BK 3N, which are 
embedded in HI streams and are among the closest companions to M81, 
is not clear. Based on the large fraction of stars with ages $\leq 200$ Myr, 
Makarova et al. (2002) argue that Ho IX and BK 3N are 
`tidal dwarfs' that formed from gas and dust that 
was pulled from M81 and/or other galaxies. While Sabbi et 
al. (2008) find older stars in the vicinity of Ho IX, they argue that these are 
tidal debris and interlopers from M81, and conclude that Ho IX is of tidal origin.
Still, Karachentshev \& Kaisin (2007) find that Ho IX lies 
along the SFR-M$_B$ relation defined by nearby galaxies. 
This is a surprising result if Ho IX is a tidal 
dwarf, as its stellar content should then differ from those of long-lived dark 
matter-dominated dwarf galaxies, with the result that its integrated $B-$band 
brightness is not a reasonable proxy for total mass when compared with these galaxies. 
For comparison, BK 3N falls well below the fiducial SFR--M$_B$ relation, as might be 
expected if its M$_B$ is elevated by a very low mass-to-light ratio with respect to 
galaxies that have a substantial fraction of their light originating from 
older populations. 

	In addition to comparatively prominent stellar ensembles like Ho IX and BK 3N, 
the debris field around M81 also contains more subtle stellar accumulations. 
Durrell et al. (2004) explore the region between M81 and NGC 3077 and find an 
ensemble of stars in the Southern Tidal Arm that has an age 30 -- 70 Myr, 
which they suggest is a tidal dwarf galaxy. de Mello et al. 
(2008) investigate stellar knots in the Arp Loop, which is located 
between M81 and M82, and find stars with ages $\leq 10$ Myr that must have formed {\it in 
situ}. Older stars are also present, and de Mello et al. argue that these formed in the disk 
of M81 and/or M82. They note that the fate of the young stellar systems is not clear; at 
one extreme they may herald the birth of a long-lived tidal dwarf galaxy, while at the 
other they may disperse, thereby forming low-density streams. Davidge (2008) finds 
a stellar structure south of M82 that contains stars with ages 
$\geq 60$ Myr. The elongated shape of this object, coupled with its low density and close 
proximity to a much larger galaxy, suggest that it will probably dissipate within a 
few hundred Myr. 

	In this paper, the wide field capabilities of the CFHT MegaCam 
are utilised to investigate the distribution of stars in the debris field in and around 
M81 and M82. A novel element of this work is that it is the first wide-field investigation of 
the debris field between these galaxies to use individual stars, as opposed to 
integrated light, to search for structures. Resolved stars can be used to 
probe areas of lower stellar density than is accessible through integrated light 
studies. Main sequence (MS) stars with masses $\geq 5$ M$_{\odot}$ 
are of particular interest for such an investigation, as 
foreground Galactic stars and background galaxies with blue colors are rare 
in the magnitude range of interest. From an astrophysical perspective, 
MS stars in this mass range have the potential of probing areas of star formation in 
the debris field that date back to the time of the interaction between M81 and M82. 

	The paper is structured as follows. The observations, reduction procedures, 
and photometric measurements are discussed in \S 2. The procedure used to probe the 
distribution of stars in the debris field is described in \S 3. Three new 
stellar groupings are identified, and the photometric properties of the stars in each 
of these are examined in \S 4. A summary and discussion of the results follows in \S 5. 
The issue of identifying individual stars in the M81 debris field from ground-based images 
is examined in the Appendix using archival HST ACS data.

\section{OBSERVATIONS \& PHOTOMETRIC MEASUREMENTS}

	The images used in this study are of a single MegaCam (Boulade 
et al. 2003) pointing, centered midway between M81 and M82. 
The detector in Megacam is a mosaic of thirty six $2048 \times 4612$ pixel$^2$ 
CCDs, that together cover an area of one degree$^2$ with 0.185 arcsec pixel$^{-1}$.
The data were recorded on the night of October 23 UT 2006. 
Four 300 sec exposures were recorded in $r'$ and $i'$,
while four 500 sec exposures were recorded in $z'$. The individual exposures were recorded 
with a square-shaped dither pattern to assist with
the identification of bad pixels and the suppression of
cosmic rays. Stars have 0.7 -- 0.8 arcsec FWHM in the final processed images, 
depending on the filter.

	The removal of instrumental signatures from the raw images was done with the 
ELIXIR processing pipeline at the CFHT. This processing included bias subtraction,
flat-fielding, and the subtraction of a fringe frame. The ELIXIR-processed images
were then aligned, stacked, and trimmed to the area that is common to all exposures.

	The photometric measurements were made with the point-spread-function
(PSF) fitting program ALLSTAR (Stetson \& Harris 1988). The PSFs, 
star catalogues, and preliminary photometric measurements 
used by ALLSTAR were obtained from routines in the DAOPHOT (Stetson
1987) package. The photometry was calibrated using
the zeropoints that are computed from standard star observations acquired 
during each MegaCam observing run, and which are placed in the data headers during 
ELIXIR processing.

	The raw photometric catalogues were culled to reject objects
with poorly determined photometric measurements. All objects in which the fit
error ($\epsilon$) computed by ALLSTAR exceeded 0.3 mag were rejected, and this
removed objects near the faint limit of the data, where the photometry is problematic. In 
addition, objects with an $\epsilon$ that is markedly higher than that of the majority 
of objects with comparable magnitude were also removed. The objects rejected
with this step tend to be either (1) non-stellar, (2) multi-pixel cosmetic defects,
and/or (3) in crowded environments. 

	Artificial star experiments were run to estimate 
sample completeness, the scatter in the photometry introduced by the combined effects of 
random noise and crowding, and the brightness at which blends of fainter stars 
account for a significant fraction of source detections. The artificial stars were
assigned brightnesses and colors that follow the main locus of bright objects in M81 and 
M82. As with actual sources, an artificial star was considered to be recovered only if
it was detected in at least two filters. The artificial star measurements were subjected
to the fit error filtering procedure that is described in the preceeding paragraph.

	As stellar density drops at a given magnitude then (1) the completeness 
fraction increases, (2) the scatter in the photometry decreases, and (3) the 
incidence of interloping blends of fainter stars decreases. 
The artificial star experiments indicate that blends tend to constitute a significant 
fraction of stars at magnitudes where the completeness fraction is $< 50\%$.
The 50\% completeness level occurs near $i' = 24.5$ throughtout most of the M81 debris 
field, although 50\% completeness occurs near $i' = 24.0$ in Ho IX, 
where the stellar density is comparatively high. 

\section{SEARCH PROCEDURE}

	The spatial distribution of objects with photometric properties that match 
those of bright main sequence (MS) stars and red supergiants (RSGs) in the M81 debris field 
are investigated in this section. Concentrations of candidate MS and RSG objects with number 
densities that exceed those outside of the M81--M82 HI debris field are identified. 
Objects on obvious stellar spiral arms associated with the main body of M81 are not 
considered. 

	The brightness and color criteria used to identify 
MS stars and RSGs were defined after inspecting the $(i', r'-i')$ CMDs of 
M81 (Davidge 2008, in preparation) and nearby control fields. Objects with 
$r'-i'$ between --0.5 and 0.0 and $i'$ between 22 and 24 were 
flagged as candidate MS stars. This region of the $(i', r'-i')$ CMD has only 
modest contamination from foreground Galactic stars and background galaxies (\S 4), 
and so objects that fall within this part of the CMD are adopted as the primary 
tracers of structure in the debris field. Objects with $r'-i'$ between 0 and 1 and 
$i'$ between 22.5 and 23.5 were flagged as candidate RSGs. Foreground Galactic stars 
and background galaxies tend to have red colors, and the region of the $(i', r'-i')$ 
CMD that monitors RSGs is prone to contamination from objects that 
do not belong to the M81 group. Consequently, candidate 
RSGs are adopted as secondary tracers of the debris field, 
as a check of detections made using MS stars. 

	There is a trade-off between angular resolution and the statistical 
robustness of source detection; if the angular resolution is too fine then 
the ability to detect objects in all but the most densely populated systems is 
hindered. Conversely, if the angular resolution is too coarse then the contrast 
between real stellar concentrations and contaminants may diminish, again complicating 
efforts to detect all but the densest stellar systems. After experimenting with 
various angular resolutions, a nominal $50 \times 50$ pixel ($9 \times 9$ arcsec) binning 
size for source counts was adopted. This corresponds to a projected spatial resolution of 
$0.15 \times 0.15$ kpc at the distance of M81. A $500 \times 500$ pixel ($90 \times 90$ 
arcsec; $1.6 \times 1.6$ kpc) binning factor was also employed to search for extremely diffuse 
stellar groupings. It should be emphasized that these large bin sizes are 
geared to detect objects that have spatial extents that greatly exceed those of 
the so-called super star clusters that have been found in interacting galaxies, 
which are possible progenitor globular clusters and have characteristic sizes of only a few 
parsecs (e.g. O'Connell 2004).

	The main criterion for identifying a stellar grouping is that it contain 
candidate MS stars with a density on the sky that exceeds that 
outside of the debris field at the $5-\sigma$ level. 
While this threshold is stringent, it lowers the chances of triggering 
detections on statistically ambiguous groupings. As an additional check, the number 
density of candidate RSGs within a projected distance of 1 kpc of the MS detection were 
also investigated. The 1 kpc radius was adopted because star formation can propogate 
spatially in stellar systems, so there may not be agreement between the locations of the 
most recent and previous areas of star formation. A 
$3-\sigma$ threshold was adopted for the red objects. The criterion that 
both candidate MS and RSGs be detected in statistically significant numbers biases 
against the discovery of extremely young systems, in which RSGs may not be present. 

	The distribution of HI, as measured by Yun et al. (1994), 
and candidate MS stars are shown in Figure 1. Not surprisingly, the dominant 
collection of MS stars is M81. While M82 is the dominant object in the northern half of the 
maps, it contains far fewer bright MS stars than M81. This is because (1) the 
angular resolution of the MegaCam data is such that individual stars could 
not be resolved in the dense central starburst of M82, with the result that 
sources in this portion of the galaxy were rejected because they had relatively high 
PSF-fitting errors values for their brightness (\S 2), and (2) 
the outer disk of M82 contains few bright main sequence 
stars (Davidge 2008, submitted to AJ). As a consequence, only 
stars on the periphery of the central star forming region of M82 are seen in Figure 1. 

	Previously identified stellar ensembles including M82 South (Davidge 2008), 
M81 West (Sun et al. 2005), and the Arp Loop are clearly seen in Figure 1. 
In addition to known objects, three heretofore unidentified groupings were discovered, 
and the central co-ordinates of these are listed in Table 1. These 
objects are referred to as Tidal Debris Objects (TDOs) 1, 2, and 3 for the remainder 
of the paper. These objects have so far eluded detection because they 
are extremely diffuse collections of stars that are distributed over 
areas spanning many square arcmin. 

	TDO 1 is 6 arcmin (5.5 kpc) from Ho IX, and 13.4 arcmin (14 kpc) from M81. It 
is not connected to Ho IX by a detectable bridge of blue stars. Still, the 
Yun et al. (1994) HI data indicates that Ho IX and TDO 1 are probably 
part of the same HI complex. TDO 1 is in the area studied by Durrell et al. (2004), and 
their Figure 1 confirms that the region near TDO 1 and TDO 2 contains a 
local over-density of blue stars, although the stellar density is not as high 
as in the cluster they find in the HI South Tidal Arm.

	TDO 2 is a pronounced concentration of blue objects immediately 
to the south and west of TDO 1. It is 15.5 arcmin (17 kpc) from the center of M81, and is 
8.5 arcmin (9 kpc) south of Ho IX. It is near the southern end of the HI complex that 
contains Ho IX.

	TDO 3 is on the major axis of M81, 16 arcmin ($\sim 17$ kpc) 
from the center of the larger galaxy, and is 5.5 
arcmin ($\sim 5.6$ kpc) from the area identified as M81 West 
by Sun et al. (2005). TDO 3 coincides with a concentration of HI emission in the 
tidal arm that also hosts M81 West (Yun et al. 1994). The central 
panel of Figure 1 shows that there are no blue stars between TDO 3 and M81 West, 
indicating that they are independent stellar concentrations. 

	Gil de Paz et al. (2007) discuss Galaxy Evolution Explorer (GALEX) images of 
nearby galaxies, including M81, and these data can be used to gain further insight into 
the new stellar groupings. More specifically, if TDO 1, 2, and 3 are relatively young 
systems then they might be sites of diffuse UV emission, such as is seen in the Arp Loop 
or Holmberg IX. A portion of the 1516\AA\ image of M81 is shown in Figure 2. 
All three of the newly discovered structures are sources of UV emission, and 
are clearly detached from the main body of M81, as expected if they are independent entities.
Moreover, the newly discovered structures have UV morphologies that are similar 
to the Arp Loop, which contains young stars (e.g. de Mello et al. 2008). The comparisons 
in Figure 2 are thus consistent with the new stellar groupings being 
concentrations of young stars. 

	In addition to the stellar groupings discussed above, the 
right hand panel of Figure 1 shows enhancements in the 
density of blue stars outside of the main body of M81, and some of these coincide with 
features in the Yun et al. (1994) HI data. Three specific enhancements 
of blue sources coincide with HI concentrations, and these are
located at ($09^{hr} 58^{min} 32.8^{sec}, +69^o 18^{'} 00^{"}$), 
($09^{hr} 52^{min} 37^{sec}, +69^o 23^{'} 23^{"}$), and ($09^{hr} 54^{min} 00^{sec}, 
+68^o 50^{'} 00^{"}$). The first of these is a band of stars that extends a few arcmin to the 
east of the Arp Loop, while the second is a complex of stars that spans 
$\sim 5 \times 5$ arcmin and is located to the north and west 
of TDO 3. The third is a spur $\sim 10$ arcmin long and 3 arcmin wide that originates 
in the main body of M81 and passes $\sim 3$ arcmin south of BK 3N. The density of 
candidate MS stars in these areas is lower than in TDO 1, 2, and 3, and the diminished 
contrast with respect to foreground stars and background galaxies 
complicates efforts to investigate the stellar contents 
of these areas. For the present study the presence of these 
features is simply noted, and their stellar contents are not investigated. 
Images with higher angular resolution would be helpful for 
distinguishing between faint galaxies and stars in these regions. 

\section{THE CMDs OF TDO 1, 2, AND 3}

\subsection{A First Look at the CMDs}

	The $(i', r'-i')$ and $(z', i'-z')$ CMDs of sources in TDO 1, 2, and 3 
are shown in the top panels of Figures 3 and 4. The CMDs of 
control fields that are located outside of the M81 -- M82 
debris field and sample the same area on the sky as the corresponding TDO 
are shown in the lower panels. Different control fields were used for each TDO, 
and the locations of the control fields are marked in Figure 1.
Aside from differences in the numbers of objects, the control field CMDs are very similar, 
suggesting that the control fields sample a common population of objects (mainly background 
galaxies), and that the distribution of objects on the CMDs are not affected by stochastic 
effects or cosmic variance.

	There is a clear over-density of blue and red objects in the TDO CMDs, 
as expected given the criteria for identifying stellar concentrations in the 
debris field (\S 3). While there are background galaxies and foreground stars 
within the boundaries of the TDOs that masquerade as RSGs, 
the $(i', r'-i')$ CMDs of the control fields are largely free of objects 
with $r'-i' < 0$ when $i' < 24$. Consequently, the brightest objects with blue colors 
in the TDO CMDs have a strong likelihood of being luminous MS stars; that these 
objects are probably point sources, as opposed to blended stellar asterisms or distant 
compact galaxies, is demonstrated in the Appendix. Finally, the $(i', r'-i')$ CMD of TDO 3 
contains a population of objects with $r'-i' > 1$ and $i' > 22$ that is not seen in the 
control fields, and these are candidate AGB stars.

\subsection{Comparisons with Isochrones}

	The CMDs of TDOs 1, 2, and 3 are compared with Z = 0.008 and Z = 0.019 isochrones 
from Girardi et al. (2004) in Figures 5 and 6. Foreground stars and background 
galaxies have been culled statistically from the CMDs by pairing 
points in each control field with those in the corresponding TDO CMD, and then deleting 
the paired objects from the TDO CMD. A maximum separation of 0.5 mag between 
points on the two CMDs was allowed, although the results are not sensitive to this value. 
The Cepheid-based distance modulus computed by Freedman et al. (1994), which is consistent 
with that determined by Bartel et al. (2007) from the expansion of the shell of SN1993J, 
has been adopted for the comparisons in Figures 5 and 6, along with 
total line-of-sight reddenings from Schlegel, Finkbeiner, \& Davis (1998). 

	All three TDOs contain relatively young stars. 
The mean color and brightness of the MS turn-off (MSTO) 
in their $(i', r'-i')$ CMDs is consistent with 
log(t) = 7.5 -- 7.7. Therefore, star formation 
in these groupings likely truncated $\sim 30 - 50$ Myr in the past. This is very similar 
to the age measured by Durrell et al. (2004) for the stellar grouping 
in the HI South Tidal Arm.

	Line blanketing affects the photometric properties of highly evolved stars 
in metal-rich and moderately metal-rich populations at wavelengths shortward of 
$\sim 1\mu$m. The decreasing impact of line blanketing 
with increasing wavelength among red stars is evident in Figures 5 and 6, in 
that at a given metallicity the upper portions of the isochrones on the 
$(z', i'-z')$ CMDs do not bend over to the same extent as those on the $(i', r'-i')$ CMDs.
The plume of bright objects with $(i'-z')_0 \sim 0.3$ in 
the $(z', i'-z')$ CMDs of TDOs 1 and 3 in Figure 6 is best matched by the Z = 0.008 models. 
The peak brightness of the RSG plume in the $(i', 
r'-i')$ CMDs of the three TDOs is also better matched by the Z = 0.008 models, which 
predict M$_{i'}$ peak magnitudes that are $\sim 0.6$ mag brighter than the Z = 0.019 
models if log(t) = 7.5.

	All three TDOs contain objects with $(r'-i')_0 \sim 0.3 - 0.4$ and M$_{i'} 
> -5$ that are RSGs with an age log(t) $\sim 8.0$. These stars are older than the 
brightest blue stars, and so it appears that TDOs 1, 2, and 3 are not simple 
stellar systems; rather, they contain stars spanning at least 0.3 dex in age. The age 
spread may be largest in TDO 3, as the red faint stars in the $(i', 
r'-i')$ CMD of this object appear to be AGB stars with an age log(t) $> 8.0$.

\subsection{Comparisons with Ho IX, M81 West, and BK 3N}

	Many of the stellar groupings in the M81 debris field share similar 
star-forming histories, with a large fraction of their stellar content forming 
in the past Gyr (e.g. Makarova et al. 2002). This is consistent with these objects 
being tidal dwarfs that condensed out of gas pulled from M81 and M82. Any ancient 
stars in the vicinity of such objects may have been pulled from the disks of the 
interacting galaxies (e.g. Sabbi et al. 2008).

	If TDO 1, 2, and 3 share a common heritage with other objects in 
the debris field then they may have similar bright stellar contents. 
The $(i', r'-i')$ and $(z', i'-z')$ CMDs of Ho IX, M81 West, and BK 3N 
are plotted in Figure 7. Ho IX is the richest stellar concentration in the M81 
debris field, and is in the same HI complex as TDO 1 and 2. 
M81 West was originally identified by Sun et al. (2005). It 
has a spectral energy distribution (SED) at visible wavelengths 
that is similar to that of the main disk of M81 (Figure 5 of Sun et al. 2005), and is 
located in the same HI tidal arm as TDO 3. The MegaCam observations indicate that 
M81 West is actually made up of a number of dense stellar groupings. Finally, 
BK 3N is a relatively compact structure, and has an integrated brightness 
that is not greatly different from that of the TDO objects.

	The peak brightnesses of the MS and RSG plumes in the CMDs of Ho IX and BK 3N 
are similar. There is a slight difference in the colors of the MS and RSG stars, in 
the sense that the Ho IX sequences are $\sim 0.1$ mag redder in $r'-i'$ than those 
in BK 3N. This is almost certainly due to a difference in line of sight reddening, 
although the Schlegel et al. (1998) maps predict almost identical reddenings. 
Still, if the Ho IX MS locus, which is overplotted on the BK 3N CMD in the upper 
row of Figure 7, is shifted to match the mean color of blue stars in BK 3N then the 
Ho IX RSG sequence matches that of BK 3N. This suggests that the RSGs in these 
systems have similar metalicities.

	Ho IX and BK 3N have high stellar densities, and so 
most of the points in their CMDs belong to these objects. 
However, contamination from foreground stars and background galaxies is more of an 
issue for M81 West. The CMDs of M81 West in Figure 7 contain sources that are 
spread over a 2 arcmin ($\sim 2$ kpc) arc on the sky, and 
contamination from foreground stars and 
background galaxies is greater than in the Ho IX and BK 3N CMDs. Still, the 
brightest object with $r'-i' < 0$ in M81 West has a comparable magnitude to those 
in Ho IX and BK 3N. As with BK 3N, the Ho IX MS and RSG sequence fall 
redward of those observed in M81 West, as expected if Ho IX has a higher 
reddening than M81 West.

	The CMDs of Ho IX, M81 West, and BK 3N are 
compared with isochrones from Girardi et al. (2002) in Figures 
8 and 9. The MS of Ho IX falls to the left of the log(t) = 7.5 isochrone in Figure 
8, as expected if the total reddening towards this system is greater than estimated 
from the Schlegel et al. (1998) maps. This being said, the 
fully de-reddened MSTO for Ho IX will be brighter than predicted by the log(t) = 7.5 
isochrone, indicating that star formation has occured within the past 30 Myr.
As for M81 West and BK 3N, the log(t) = 7.5 isochrones match the color of the MS 
in both objects, although the MSTO of BK 3N may be slightly brighter than predicted by the 
log(t) = 7.5 isochrones. It is also perhaps interesting that the $(i', r'-i')$ CMDs of 
BK 3N do not contain objects with ages log(t) $> 8.0$. Makarova et al. (2002) 
find that the RGB stars in the vicinity of BK 3N are uniformly distributed on the sky, 
indicating that they probably belong to M81.

	The RSG sequences on the $(z', i'-z')$ CMDs of all three systems 
are matched best by the Z = 0.008 isochrones, and 
an independent confirmation of this metallicity estimate comes from 
the shape of the RSG sequence on the $(i', r'-i')$ CMD. 
The red plume on the $(i', r'-i')$ CMD of Ho IX is roughly 
vertical, whereas the isochrones suggest that if the stars had a solar metallicity 
then they would bend over at the bright end. Sabbi et al. (2008) find that Z = 
0.008 isochrones also give the best match to their ACS observations of Ho IX, and 
Makarova et al. (2002) report that the chemical abundance of an HII region in Ho IX is 
consistent with Z = 0.008.

	In summary, the comparisons in Figures 8 and 9 suggest that the stars that formed 
$\sim 30$ Myr in the past in Ho IX, M81 West, and BK 3N have Z = 0.008, which is in good 
agreement with what is seen in TDO 1, 2, and 3. The most recent episodes of 
star formation in Ho IX, M81 West, and BK 3N all occured some 30 Myr in the past, also 
in agreement with what is seen in TDO 1, 2, and 3. Finally, 
the CMDs of M81 West and TDO 3 share another similarity, in that both contain a 
population of stars with M$_{i'} > -5$ and $r'-i' < -1$, which the isochrones suggest 
may be stars that have ages approaching 1 Gyr.

\subsection{The Integrated Photometric Properties of TDO 1, 2, and 3}

	The surface brightness and total brightness of TDO 1, 2, and 3 can be estimated 
if it is assumed that their stellar contents are the same as in Ho IX, and then 
scaling star counts from Ho IX to those in the TDO objects. This is done here 
using the number of sources with $r'-i'$ between 0 and --0.5 and with $i'$ between 
23 and 22, which is a region of the $(i', r'-i')$ CMD where contamination from 
objects that do not belong to the M81 group is modest. 
There are 44 stars in this portion of the Ho IX CMD, whereas there 
are 4 such stars each in of the TDO CMDs. If the integrated brightness of Ho IX is 
$V = 13.9$ and $K = 11.6$ (Dale et al. 2007), then the TDOs have 
integrated magnitudes $V \sim 16.5$ and $K \sim 14.2$. A total area of 43000 
arcsec$^2$ was probed in TDO 1, and so the mean surface brightness is 28.1 mag 
arcsec$^{-2}$ in $V$ and 25.8 mag arcsec$^{-2}$ in $K$. As for TDO 2, 
an area of 10500 arcsec$^2$ was studied, so that the mean surface brightness is 
26.6 mag arcsec$^{-2}$ in $V$, and 24.3 mag arcsec$^{-2}$ in $K$. Finally, 25000 
arcsec$^{2}$ was investigated in TDO 3, so the mean surface brightness is 
27.5 mag arcsec$^{-2}$ in $V$ and 25.2 mag arcsec$^{-2}$ in $K$.
 
\section{DISCUSSION \& SUMMARY}

	Images obtained with the CFHT MegaCam have been used to investigate the 
distribution of bright stars in the M81 debris field. 
A novel aspect of this work is that individual stars, rather than 
integrated light, are used to search for structure. Such a search is made possible by the 
large MegaCam science field, combined with the sub-arcsec angular resolution of these data. 
A significant benefit of using individual stars is that objects 
that have photometric properties that are relatively common among stars in the M81 
group, but are rare among foreground Galactic stars and background galaxies, 
can be selected to enhance sensitivity. An unfortunate drawback is that the current search 
is restricted to intrinsically bright, inherently rare, stars. Deeper studies will 
allow structure in the debris field to be better defined.

	The search uses sources with blue colors in a magnitude 
range that contains massive MS stars and blue supergiants in the M81 group. 
This was supplemented with a secondary search geared towards detecting objects with 
colors and magnitudes that are appropriate for RSGs. While rare, background galaxies 
with colors that are similar to those of massive MS stars are still present, 
and galaxy clusters that contain a mix of unreddened 
star-forming galaxies and redder galaxies are one source of 
contamination in the present search. In fact, two clumps of blue 
and red sources that are made up of obviously extended objects were found 
that met the threshold criteria for detection.

	Bright main sequence stars are found throughout the M81 debris field, 
and three previously unidentified young stellar concentrations are 
discovered. The area considered here does not include the candidate tidal dwarf 
identified by Durrell et al. (2004) in the HI South Tidal Arm. Still, the three objects found 
here are similar to that cluster, both in terms of spatial extent ($\sim 1$ kpc) and age.

\subsection{The Past and Future of TDO 1, 2, and 3}

	The distribution of star clusters on the 
$(V-I, B-V)$ diagram suggests that there was a period 
of elevated star formation throughout the M81 disk 
a few tens of Myr in the past (Figure 3 of Chandar, 
Tsvetanov, \& Ford 2001). This may also have been the case in the tidal debris field. 
The ages estimated for the brightest MS and RSG stars indicate that TDO 1, 2, and 3, as 
well as other objects such as Ho IX, have been forming stars 
for at least some fraction of the past $\sim 100$ Myr. 
The time interval over which TDO 1, 2, and 3 have been forming stars 
is the typical timescale for elevated levels of star-forming activity in starburst 
galaxies (e.g. Marcillac et al. 2006).

	The young stars in the TDOs could have formed {\it in situ}, 
as the gas density exceeds the Kennicutt (1989) threshold. 
Using the Adler \& Westpfahl (1996) rotation curve for M81, 
the threshold gas density for star formation, $\Sigma_{c}$, near TDO 1 and 2 is 
$\Sigma_{c} \sim 3.4$ M$_{\odot}$ pc$^{-2}$, while for TDO 3 $\Sigma_{c} \sim 4.5$ 
M$_{\odot}$ pc$^{-2}$. This corresponds to HI densities $\sim 3 \times 10^{20}$ cm$^{-2}$ 
for TDO 1 and 2, and $4 \times 10^{20}$ cm$^{-2}$ for TDO 3. The observed HI densities 
in the vicinity of the TDOs comfortably exceed these values (e.g. Yun et al. 1994; Adler 
\& Westpfahl 1996). 

	The newly discovered systems contain MS and and RSG stars that, 
based on comparisons with isochrones, have different ages; hence, they appear to 
be composite systems, as opposed to simple stellar systems. 
However, could the RSG sequences in the TDO CMDs be part of a stellar 
system that is not associated with the MS stars, such as a diffuse population of RSGs 
pulled from the M81 disk? The answer to this question is a qualified `no'. 
A requirement in the search procedure employed here is that stellar concentrations show up 
as statististically significant groupings of both blue and red objects. 
A caveat is that offsets between MS and RSG groupings on the order of kpc and smaller 
scales were tolerated, since areas of star formation can propogate spatially. Therefore, 
the RSGs are not part of a diffuse population that has been pulled from M81 or M82. Still, 
there is no evidence other than a small projected separation that the MS and RSG 
components are physically related, and hence are part of a single `system'.

	Mirabel, Dottori, \& Lutz (1992) find a dwarf galaxy-like structure  
in a tidal tail emerging from NGC 4038/39, and the formation of such `tidal dwarfs' 
is predicted by simulations of galaxy-galaxy interactions (e.g. Barnes \& Hernquist 1992). 
Tidal dwarfs form from gas-rich debris and have only modest dark matter contents 
when compared with traditional galaxies that formed in the heart of dark matter 
halos (Barnes \& Hernquist 1992; Wetzstein, Naab, \& Burkert 2007). 
Simulations run by Cox et al. (2006) show that fragments from tidal interactions may 
last only a few hundred Myr. 

	Despite not having a large dark matter component, 
tidal dwarfs may not be disrupted quickly by feedback, 
and Recchi et al. (2007) find that star formation may continue for hundreds of Myr, 
or until the system is destroyed by tidal forces. In fact, the majority of objects that 
form in debris fields are probably not long-lived, 
as there appears to be only a limited region of parameter 
space in which tidal dwarfs can survive for even a fraction of the Hubble time.
Bournaud \& Duc (2006) find that only 10 -- 20\% of tidal 
dwarfs survive up to 10 Gyr, and even then only if the interaction geometries 
are favourable. Many of the dwarfs formed in their simulations disperse within 
a few hundred Myr, and the longest-lived tidal structures are those in which the orbits 
of the interacting galaxies are coplanar to within 40$^o$, so that tidal 
features form close to the disk planes. The longest lived tidal dwarfs in the 
Bournard \& Duc (2006) simulations have masses $\geq 10^8$ M$_{\odot}$.
Metz \& Kroupa (2007) find that systems with masses 10$^7$ M$_{odot}$ may also 
survive, albeit as low surface brightness dwarf spheroidals (see also Kroupa 1998). 

	The locations of TDO 1, 2, and 3 in the HI debris field, coupled with 
the disk-like metallicities of the brightest stars and the density of gas in the 
regions around these objects, which exceeds that necessary to trigger star formation 
(see above), suggests that they may have formed 
from tidal debris. TDO 1, 2, and 3 may thus be considered to be tidal dwarfs, although the 
reader should keep in mind that they are probably destined to be -- at least by 
cosmological standards -- short-lived structures, surviving less than $\sim 1$ Gyr. 
With M$_V \sim -11$ (\S 4), then TDO 1, 2, and 3 have masses $\sim 10^5 - 
10^6$ M$_{\odot}$, and so are less massive than the long-lived tidal dwarfs that have 
been found in simulations to survive for long periods. 
Lacking knowledge of the orbital paths of the TDO objects with respect to M81, 
and the kinematics of individual stars within these systems,
it can not be said with certainty whether or not they are dissipating. Still, 
assuming a M/L ratio = 0.2, which is appropriate for an age 
of 100 Myr (Mouhcine \& Lancon 2003), and a total mass of $4 \times 10^{11}$ 
M$_\odot$ for M81 (Schroder et al. 2002) then the tidal radii 
of the TDOs fall between 100 and 200 kpc. This is 
significantly larger than their projected separations from M81, 
suggesting that they are prone to disruption. Streams of material originating 
from TDO 1, 2, and 3 would be one signature of this happening. 
While there is no evidence in the MegaCam data for stellar streams, 
such features may be too diffuse to detect without going significantly fainter. 

\subsection{Other Objects in the Debris Field}

	A number of stellar groupings have previously been identified in the M81 debris field, 
and the star-forming histories of these are of interest for understanding not only the 
evolution of the TDO objects, but also of the debris field 
in general. Many of the previously discovered objects fall inside the 
MegaCam science field, and so it is possible to compare their CMDs with those of the TDO 
objects. The CMDs of all three TDOs are similar to those of Ho IX, BK 3N, and 
M81 West at the bright end (\S 4), indicating similarities in their 
recent star-forming histories and mean metallicities. These similarities suggest that 
star formation throughout the tidal debris field was triggered by common events, 
presumably related to the encounter between M81 and M82. 
In the case of Ho IX, TDO 1, and TDO 2, the similarity in stellar content 
and metallicity is perhaps not surprising, given their close proximity. 
In fact, that they are located in the same HI complex suggests that these systems 
probably formed in the same tidal arm. 

	Karachentsev \& Kaisin (2007) use the integrated properties 
of Ho IX and BK 3N to investigate their evolution. 
With the caveat that the HI mass associated with a stellar system in the 
debris field is hard to determine with confidence since some (or all) of the HI along the 
line of sight may not belong physically to the object in question, Karachentsev \& Kaisin 
(2007) estimate the rate of consumption of HI in a number of M81 group galaxies based on 
the present SFR, and construct a diagram that compares the past and projected fuel 
consumption rates normalized to the available gas. The position of Ho IX and BK 3N on this 
diagram suggests that they have enough gas to fuel star formation for some time to come 
if they maintain their present SFR. For comparison, the location 
of M82 and many M81-group spirals and dwarf irregulars on 
this diagram suggest that they will deplete their gas reservoirs within a Hubble time if 
they continue to form stars at their current rates.

	The rosey prognosis for sustainable star formation in Ho IX
is not a consequence of a low SFR. Indeed, Karachentsev \& Kaisin (2007) find that 
Ho IX falls on the SFR-M$_B$ relation defined by nearby galaxies. 
If, as suggested by photometric studies, Ho IX is a young stellar system, then 
adjusting its integrated brightness to account for its age when compared with 
traditional long-lived dark matter-dominated dwarf galaxies will shift it to a region 
of the SFR-M$_B$ diagram that falls above the fiducial relation. 
That a large gas reservoir is associated with Ho IX is qualitatively consistent with the 
gas-rich nature of tidal dwarfs predicted by Wetzenstein et al. (2007).
That Ho IX, BK 3N, M81 West, and the three TDO systems
have similar metallicities, despite differences in their integrated brightness, is 
also consistent with a tidal dwarf origin.

\subsection{Broader Implications}

	The M81 debris field may provide clues into the origin of stars that 
are seen in the extraplanar regions of other galaxies. Mouhcine (2006) investigates the 
metallicity distribution functions (MDFs) of stars in the outer regions 
of nearby spiral galaxies. The MDFs contain a tail of 
metal-poor stars that he suggests may have originated in dwarf 
galaxies or protogalactic fragments that formed either outside of the influence of the 
dark matter halo of the host spirals, or before the dominant dark matter halo was assembled. 
The characteristic metallicity of the extraplanar component 
also is seen to increase with galaxy luminosity, suggesting that -- unlike the metal-poor 
stars -- the more metal-rich stars in the MDF formed within the dark matter halo of the 
dominant galaxy, as their chemical enrichment was evidently influenced by 
the depth of the potential well of the host spiral. 

	Mouhcine (2006) suggests that the metal-rich stars in the MDF of the 
target galaxies originated in globular clusters. However, could these stars have 
come from the disk, as a result of tidal stripping or heating due to tidal interactions? 
If, as argued by Hammer et al. (2007), the majority of spiral galaxies 
have experienced mergers during the past $\sim 6$ Gyr 
then the tidal stirring of disks is not a rare event. Some of 
the galaxies in the Mouhcine sample are viewed almost edge-on, and 
Figure 1 of Mouchine (2005) indicates that in some cases 
the sampled fields extend only a few vertical disk scale lengths off of the disk plane. 
In fact, the fields observed by Mouhcine (2006) probe extraplanar distances that are 
similar to those where metal-rich stars are prevalent in M31; the spatial distribution 
and kinematic properties of these stars indicate that they are not part of the 
classical halo in M31 (Kalirai et al. 2006; Chapman et al. 2006). The spatial distribution of 
the metal-rich stars in extragalactic spirals will provide clues as to their origin. 
If the metal-rich component defines a flattened system then 
this would be consistent with a disk origin. In contrast, a 
centrally concentrated spheroidal or near-spheroidal distribution would be more 
consistent with an origin tied to the bulge and/or its associated globular clusters. 

\acknowledgements{Sincere thanks are extended to the anonymous reviewer, who 
suggested changes that greatly improved the paper.}

\appendix

\section{The Nature of the Blue Sources in the M81 Debris Field}

	Blending is an obvious consideration when observing intrinsically bright stars 
in extragalactic systems. In the M81 group, one arcsec corresponds to 
a projected distance of 17.6 parsec, and so
numerous stars inevitably fall within the MegaCam seeing disk. 
This being said, the nature of the stellar luminosity function 
is such that the vast majority of these objects are intrinsically faint, 
so that the impact on the photometry of bright stars is often insignificant.
In the current study, the impact of contamination from other stars within the 
seeing disk is mitigated somewhat by the relatively low 
stellar density in the M81 debris field. That the ages estimated for 
Ho IX and BK 3N from the brightest main sequence stars and RSGs are 
consistent with those obtained from HST data suggests that blending 
is not an issue. Indeed, the stellar densities in TDO 1, 2, and 3 are lower than in Ho IX and 
BK 3N, and so the incidence of blending will be even lower in these systems.

	Could the blue sources that are the basis of the search for stellar 
groupings be stellar asterisms or background galaxies? This question can 
be answered by examining the blue sources in HST images. Here, we consider 
deep F814W ACS images of the southern portion of the Arp Loop, which were recorded 
as part of program 10915 (PI J. Dacanton). These data are well suited 
for this task because (1) the Arp Loop contains a number of blue sources in a relatively 
compact area, and (2) they were recorded with moderately long exposure times, thereby 
assisting in the identification of faint galaxies.

	Six blue stars were detected by MegaCam in a $1 \times 1$ 
arcmin$^2$ region that is located midway between the FUV sources 6 and $7+8$ 
discussed by de Mello et al. (2008). A $2 \times 2$ arcsec section of the 
ACS F814W image, centered on each of these stars, is shown in Figure A1. In each case 
there is a dominant central point source, which is the star that was 
identified by MegaCam. At the angular resolution of the ACS data, the central stars 
appear to be point sources, and hence are not compact star clusters or 
obvious background galaxies.

	Four of the six stars have an obvious companion within $\sim 1$ arcsec, 
and these companions also appear to be point sources. 
The presence of such companions is perhaps not surprising 
given that the Arp Loop is, at least by the standards of the M81 debris 
field, a comparatively high density environment. This being said, the companions 
are much fainter than the central star, with the smallest magnitude difference between 
the central source and the companion being 1.7 magnitudes. Moreover, only in one case 
does the companion fall within the seeing disk (FWHM $= 0.75$ arcsec) of the MegaCam data.
The ACS data are thus consistent with the blue sources in the MegaCam 
data being intrinsically bright main sequence stars.

\clearpage

\begin{table*}
\begin{center}
\begin{tabular}{ccc}
\tableline\tableline
TDO \# & RA & Dec \\
 & (E2000) & (E2000) \\
\tableline
1 & 09:58:17 & $+68:57:40$ \\
2 & 09:58:02 & $+68:54:10$ \\
3 & 09:54:05 & $+69:18:49$ \\
\tableline
\end{tabular}
\caption{Co-ordinates of the Three New Stellar Groupings}
\end{center}
\end{table*}

\clearpage
\begin{figure}
\figurenum{1}
\epsscale{0.85}
\plotone{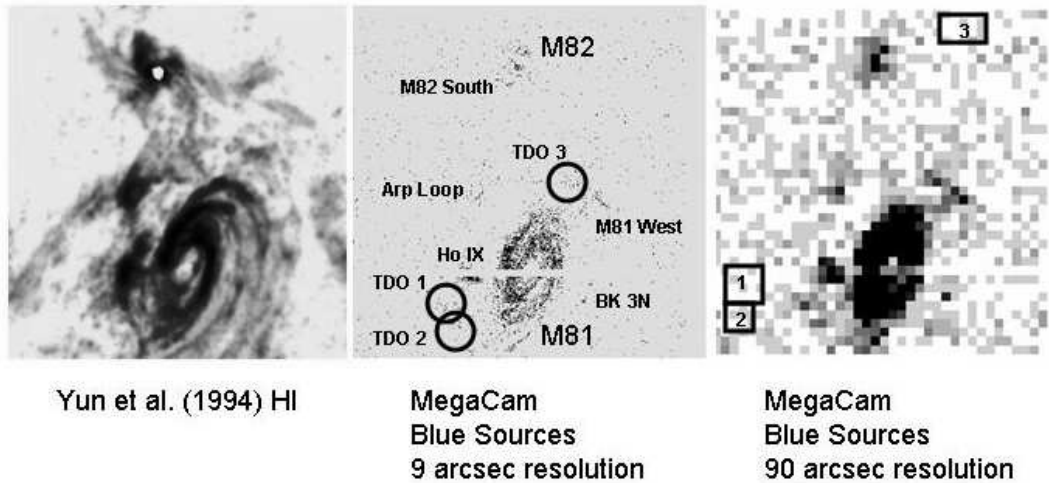}
\caption
{The distribution of blue sources that are candidate main sequence stars is 
shown in the middle ($9 \times 9$ arcsec binning) and right hand 
($90 \times 90$ arcsec binning) panels. HI emission, as mapped by Yun et al. (1994), is shown 
in the left hand panel. The spatial scale and orientation of the HI map matches those 
of the MegaCam data, with North at the top and 
East to the left. TDO 1, 2, and 3 are identified, as are 
other sources in the debris field that have been 
previously discovered. The locations of the control fields that are used in the analysis 
of the CMDs of TDO 1, 2, and 3 in \S 4 are indicated on the 90 arcsec resolution map. 
Note that diffuse ensembles of candidate MS stars are seen 
in the $90 \times 90$ arcsec map that extend to the east of the 
Arp Loop, to the north and west of TDO 3, and to the south of BK 3N, 
and these coincide with areas of HI emission.}
\end{figure}

\clearpage
\begin{figure}
\figurenum{2}
\epsscale{0.85}
\plotone{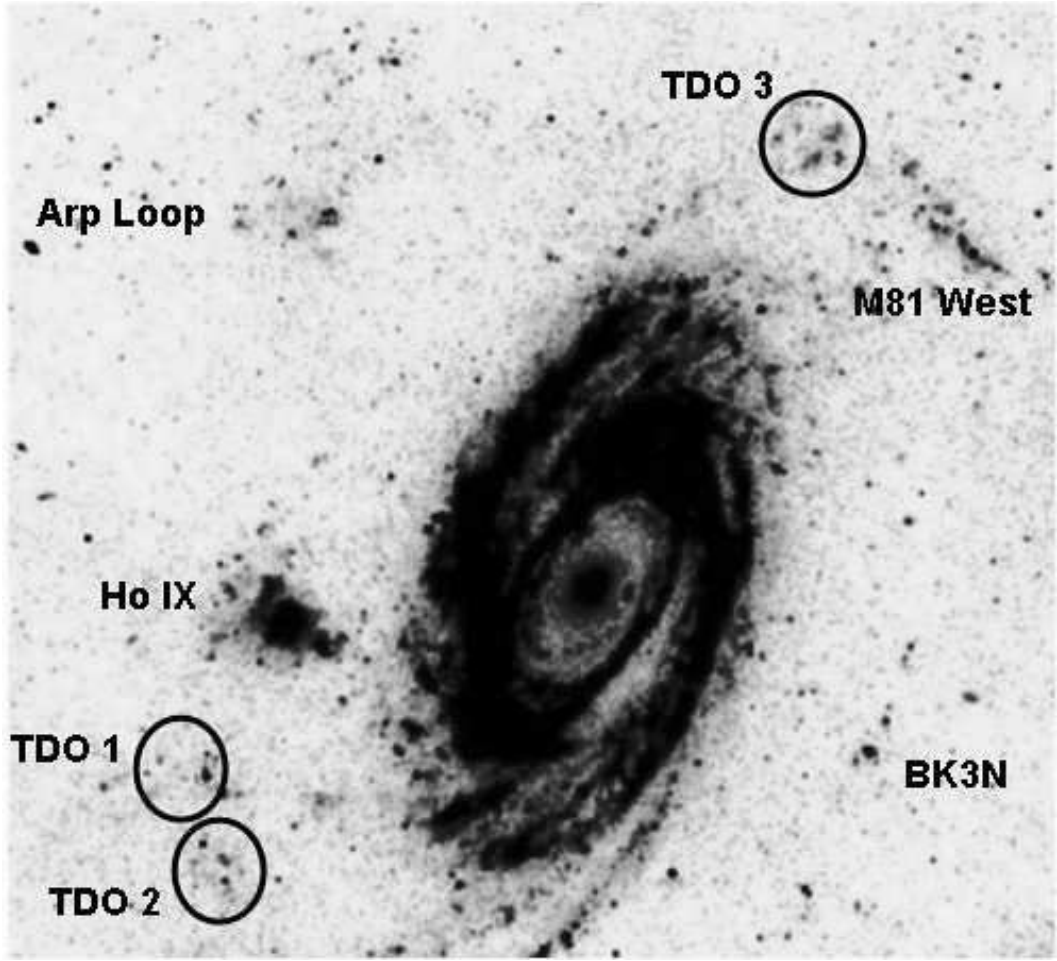}
\caption
{A section of the 1516\AA\ image of M81 from Gil de Paz et al. (2007). The locations of 
the newly discovered stellar groupings are indicated, and previously identified 
structures are marked. Note that TDOs 1, 2, and 3 are sites of extended far-ultraviolet 
emission and have an appearance that is similar to the Arp Loop. This is consistent with these 
objects being relatively young stellar systems.}
\end{figure}

\clearpage
\begin{figure}
\figurenum{3}
\epsscale{0.85}
\plotone{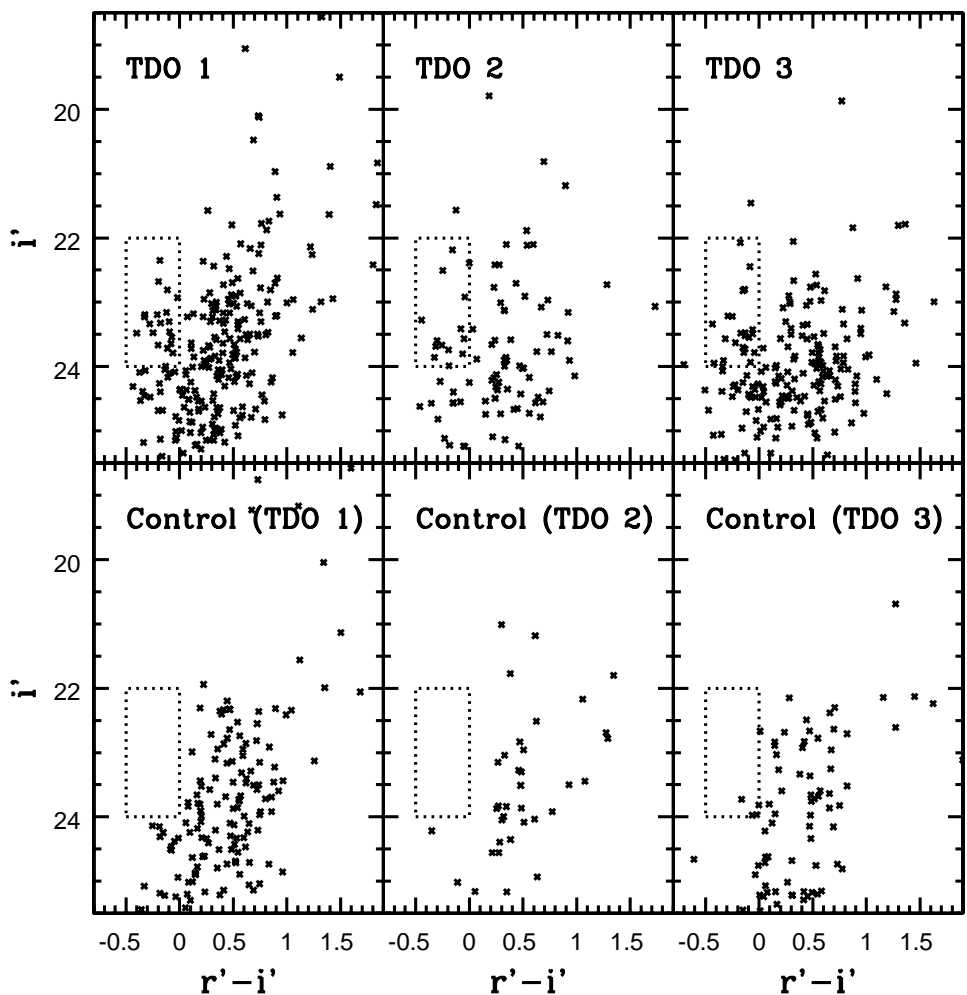}
\caption
{The $(i', r'-i')$ CMDs of TDO 1, 2, and 3 and their control fields. 
The dotted lines mark the area in the CMDs that is assumed to contain
bright main sequence stars in the search for stellar groupings in the debris field. 
There is a clear excess population of blue objects in the CMDs of all three TDOs 
when compared with the CMDs of the control fields.}
\end{figure}

\clearpage
\begin{figure}
\figurenum{4}
\epsscale{0.85}
\plotone{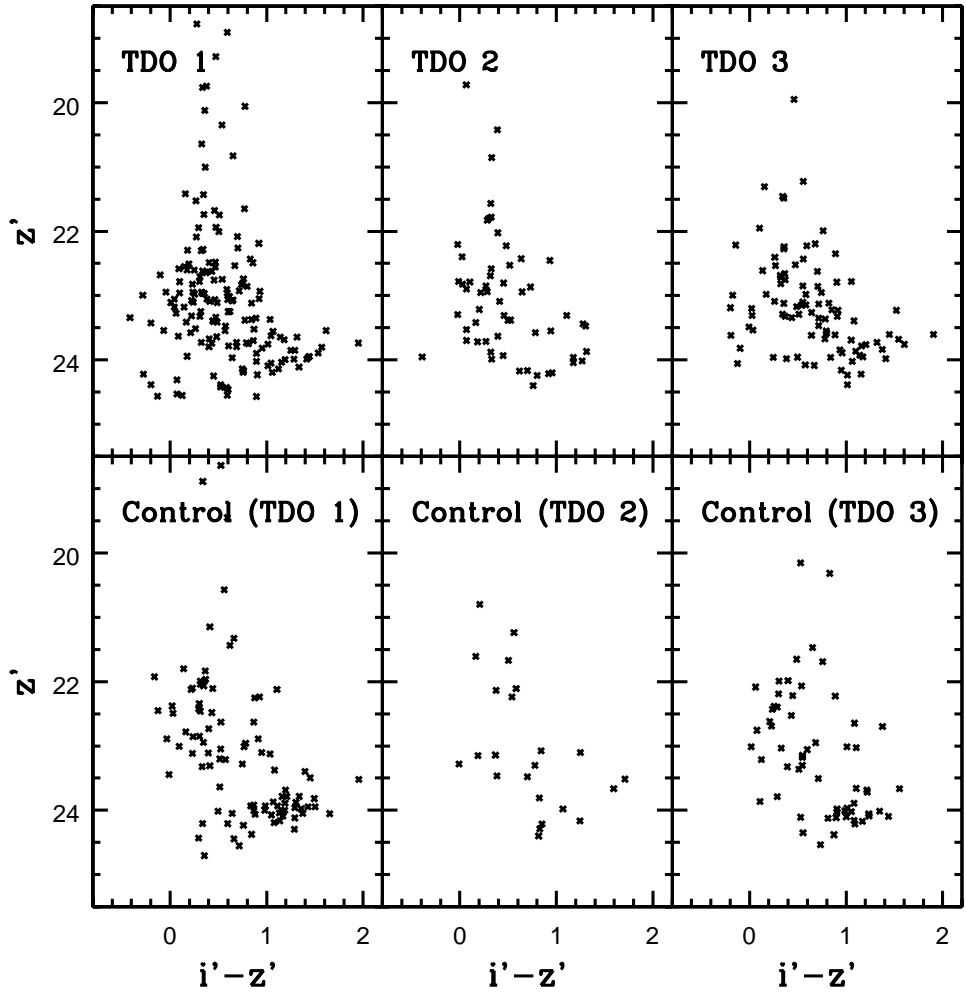}
\caption
{The the $(z', i'-z')$ CMDs of TDO 1, 2, and 3 and their control fields.}
\end{figure}

\clearpage
\begin{figure}
\figurenum{5}
\epsscale{0.85}
\plotone{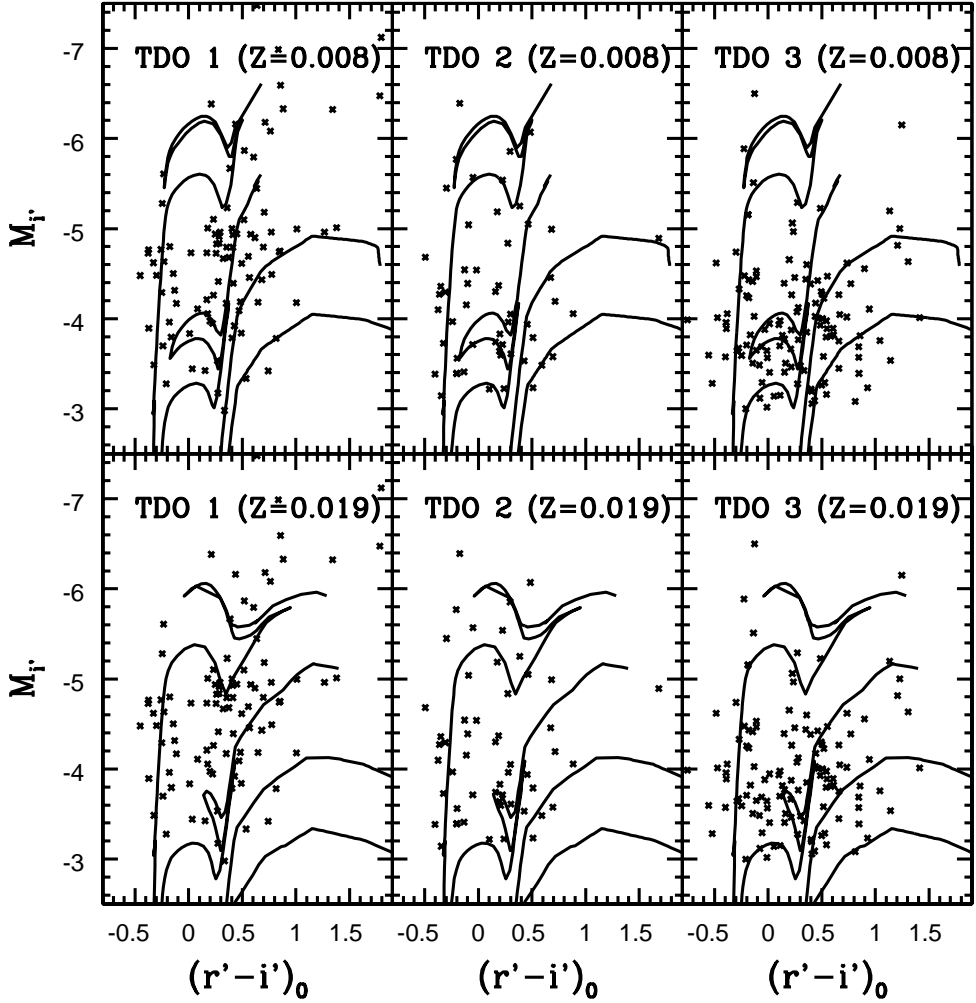}
\caption
{The $(M_{i'}, r'-i')$ CMDs of TDOs 1, 2, and 3 are compared with Z = 0.008 and 
Z = 0.019 isochrones from Girardi et al. (2004). The isochrones have ages 
log(t$_{yr}$) = 7.5, 8.0, 8.5, and 9.0. The CMDs have been cleaned of foreground stars 
and background galaxies using the statistical procedure described in the text. 
The Freedman et al. (1994) distance modulus of 27.8 is adopted for this 
comparison, as is the Schlegel et al. (1998) reddening estimate for each object.} 
\end{figure}

\clearpage
\begin{figure}
\figurenum{6}
\epsscale{0.85}
\plotone{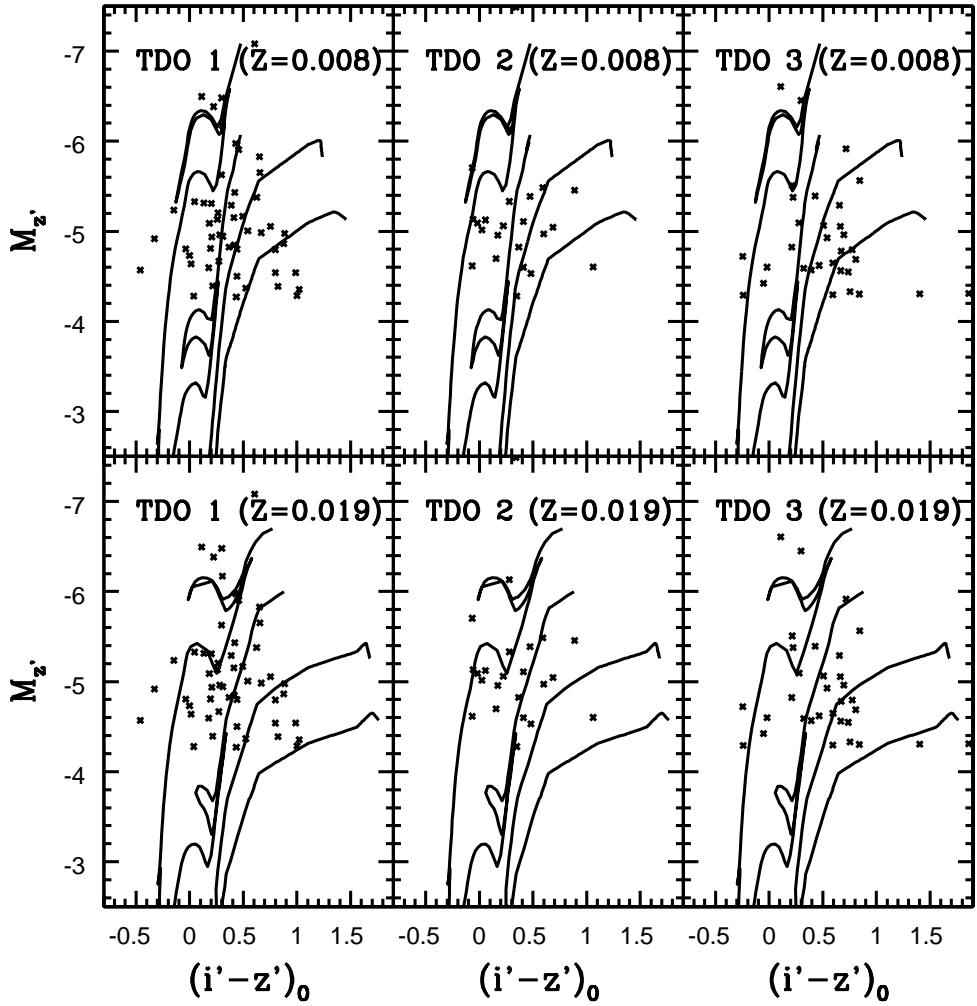}
\caption
{The same as Figure 5, but showing the $(M_{z'}, i'-z')$ CMDs of TDOs 1, 2, and 3.} 
\end{figure}

\clearpage
\begin{figure}
\figurenum{7}
\epsscale{0.85}
\plotone{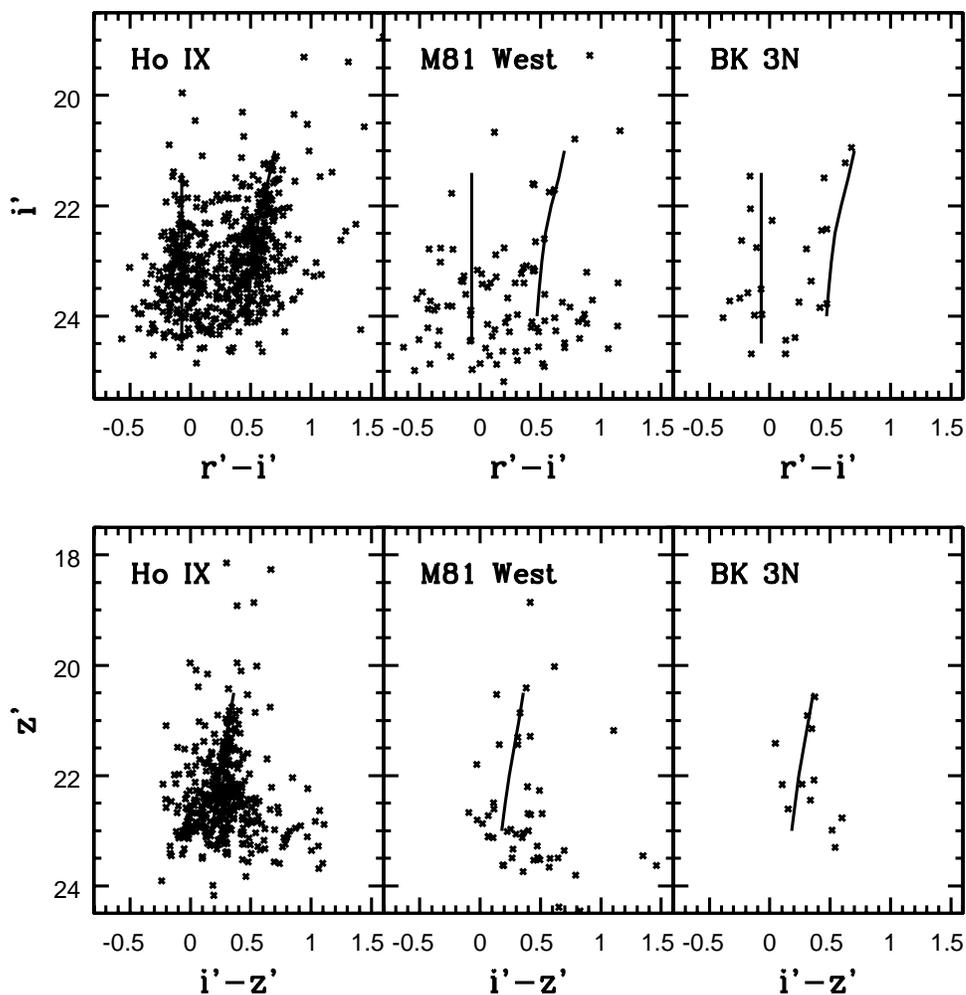}
\caption
{The $(i', r'-i')$ and $(z', i'-z')$ CMDs of Ho IX, M81 West, and BK 3N. The lines show 
the loci of RSGs and main sequence stars in Ho IX. Note that the Ho IX MS ridgeline is 
offset from those in M81 West and BK 3N, suggesting a higher total reddening 
for Ho IX than in the other systems.}
\end{figure}

\clearpage
\begin{figure}
\figurenum{8}
\epsscale{0.85}
\plotone{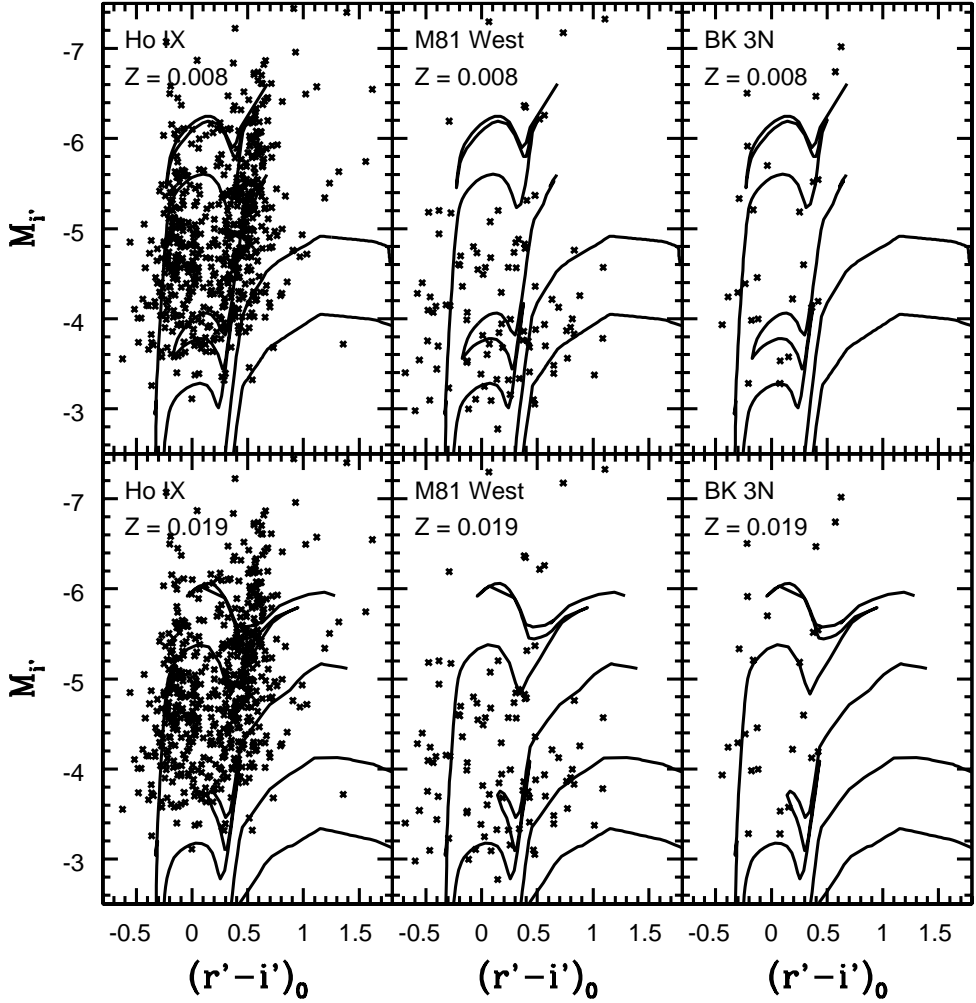}
\caption{The $(M_{i'}, r'-i')$ CMDs of Ho IX, M81 West, and BK 3N are compared with Z = 
0.008 and Z = 0.019 isochrones from Girardi et al. (2004). The isochrones have ages 
log(t) = 7.5, 8.0, 8.5, and 9.0. A distance modulus of 27.8 (Freedman et al. 1994) 
is assumed, as are total reddenings from Schlegel et al. (1998). Based on 
the color of the MS, it appears that the reddening towards Ho IX may be underestimated. 
The shape of the RSG plume in these galaxies favours Z = 0.008, rather than Z = 0.019.}
\end{figure}

\clearpage
\begin{figure}
\figurenum{9}
\epsscale{0.85}
\plotone{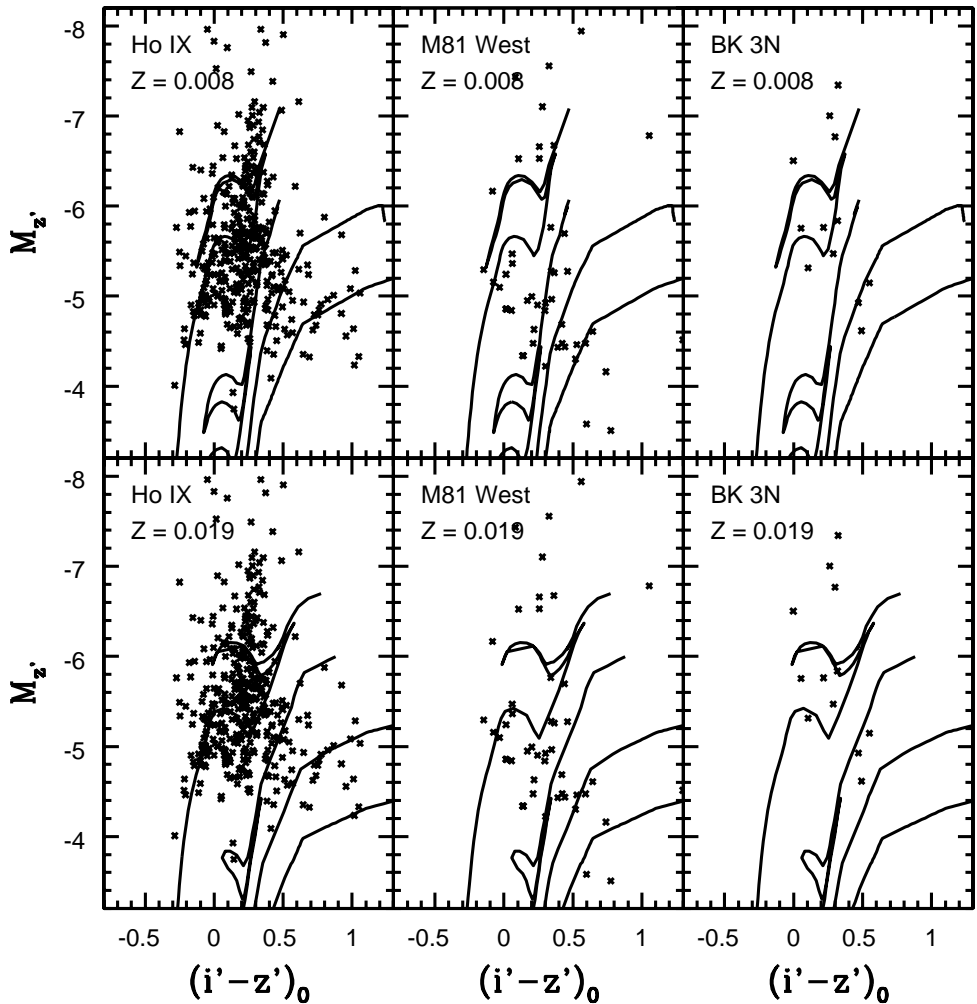}
\caption{The same as Figure 8, but showing $(M_{z'}, i'-z')$ CMDs. Note that the Z=0.008 
isochrones better match the color of the RSG plume in all three galaxies than the 
Z=0.019 isochrones.} 
\end{figure}

\clearpage
\begin{figure}
\figurenum{A1}
\epsscale{0.85}
\plotone{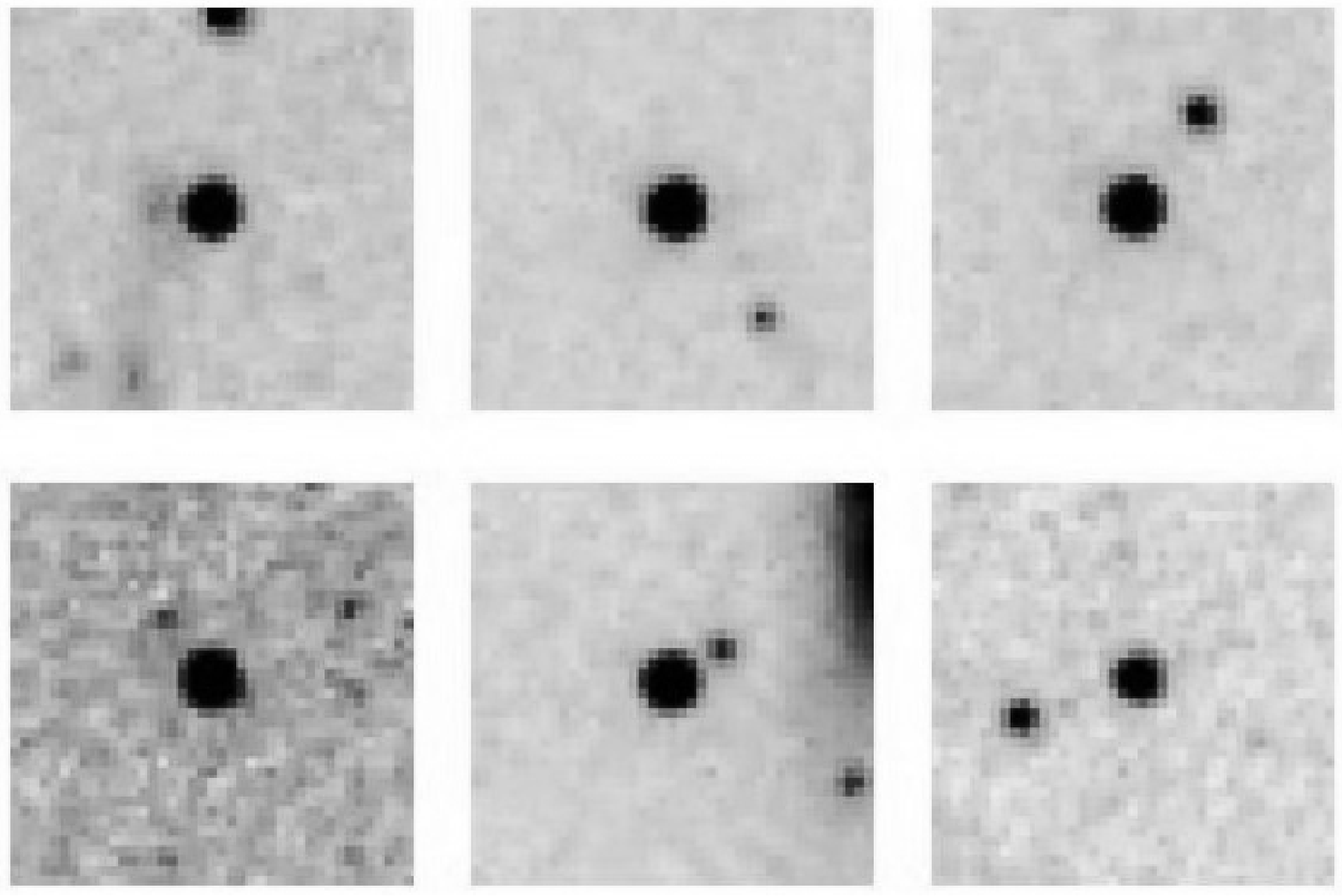}
\caption{ACS F814W images of stars that have blue colors in the MegaCam data. 
Each frame subtends $2 \times 2$ arcsec. The dominant object 
in each frame is a point source, as opposed to a stellar asterism or galaxy. The 
brightest companion is five times fainter than the central star, and in 
only one case does the companion fall within the 
seeing disk of the MegaCam data. These ACS data support the notion that the 
blue sources used to map structure in the M81 debris field are main sequence stars in the 
M81 group.}
\end{figure}

\end{document}